\documentclass[amsmath,prl,aps,twocolumn,showpacs,superscriptaddress,floatfix]{revtex4}

\usepackage{amsfonts}
\usepackage{bm}
\usepackage{graphicx}
\usepackage{color}

\def\hmsun{\ {\rm M_\odot/{\it h}}}

\def\hmmpc{\ {\rm {\it h}Mpc^{-1}}}

\def\ln{{\rm ln}}

\def\vk{\mathrm{\bf k}}

\def\fnl{f_{\rm NL}}
\def\gnl{g_{\rm NL}}

\def\apjl{Astrophys.\ J.\ Lett.}
\def\mnras{Mon.\ Not.\ R.\ Astron.\ Soc.}

\def\apj{Astrophys.\ J.}

\def\apjs{Astrophys.\ J. Supp.}

\def\prd{Phys.\ Rev.\ D}

\def\jcap{{JCAP\ }}

\newcommand{\be}{\begin{equation}}
\newcommand{\ee}{\end{equation}}
\newcommand{\bea}{\begin{eqnarray}}
\newcommand{\eea}{\end{eqnarray}}
\def\ba#1\ea{\begin{align}#1\end{align}}
\newcommand{\D}{\Delta}
\newcommand{\rhob}{\overline{\rho}}

\newcommand{\s}{\sigma}
\renewcommand{\d}{\delta}
\renewcommand{\a}{\alpha}

\newcommand{\Sk}{S^{(3)}}

\newcommand{\<}{\langle}
\renewcommand{\>}{\rangle}
\newcommand{\bo}{{\rm b}_{\rm I}}

\newcommand{\iMpch}{\,h/{\rm Mpc}}

\newcommand{\refeq}[1]{Eq.~(\ref{eq:#1})}

\newcommand{\reffig}[1]{Fig.~\ref{fig:#1}}

\def\bm#1{\textbf{\em #1}}
\def\Mm{\mathcal{M}}

\def\Pp{P_\phi}

\definecolor{RedWine}{rgb}{0.743,0,0}
\definecolor{RoyalBlue}{rgb}{0.25,.41,.88}
\definecolor{ForestGreen}{rgb}{.13,.54,.13}
\definecolor{DeepPurple}{rgb}{.72,.18,1}

\begin{document}

\title{
Accurate Predictions for the Scale-Dependent Galaxy Bias from Primordial
Non-Gaussianity
}
\author{Vincent Desjacques}
\affiliation{Institute for Theoretical Physics, University of Z\"urich,
Winterthurerstrasse 190, CH-8057 Z\"urich, Switzerland}
\author{Donghui Jeong}
\affiliation{
California Institute of Technology, Mail Code 350-17, Pasadena, CA, 91125
}
\author{Fabian Schmidt}
\affiliation{
California Institute of Technology, Mail Code 350-17, Pasadena, CA, 91125
}
\date{\today}


\begin{abstract}
The large-scale clustering of galaxies can serve as a probe of 
primordial non-Gaussianity in the Universe competitive with the 
anisotropies of the CMB. Here, we present improved theoretical predictions 
which include an important, previously overlooked correction to the bias.  
We demonstrate that the new predictions are able to reproduce the 
results of N-body simulations, explaining the significant departures seen 
from previous theoretical results. These refined predictions open the way to 
accurate constraints on primordial physics with large-scale structure surveys.
\end{abstract}


\maketitle


Measurements of the temperature and polarization anisotropies of the 
cosmic microwave background (CMB) radiation strongly support inflation 
as the mechanism generating the initial curvature perturbations 
\cite{wmap7}, here parametrized by the Bardeen potential perturbations 
$\phi = \Phi_H$ \cite{bardeen}.  
While the simplest, single-field slow-roll models predict nearly Gaussian 
fluctuations, many other inflationary scenarios yield a potentially 
detectable level of primordial non-Gaussianity, which can generally 
be described in terms of the higher-order ($N\geq 3$) correlation functions 
of $\phi$ in Fourier space, $\xi_\phi^{(N)}$.  
Because different seeding mechanisms of the initial curvature perturbations 
leave distinct signatures in the shape and amplitude of these higher-order 
correlation functions, a precise measurement of non-Gaussianity would 
strongly constrain the physics of inflation. 

As the primordial curvature perturbations eventually grow into the large 
scale structure observed today, their higher order correlation functions 
can be determined through a measurement of the clustering of dark matter 
halos \cite{ds10review} which host the galaxies. For example, 
primordial non-Gaussianity of the local $\fnl\phi^2$-type induces a strongly 
scale-dependent bias $\D\bo(k,z)$ in the clustering of halos \cite{ddhs08}.  
The power spectrum $P_h$ of halos of mass $M$ at redshift $z$ is related 
to the matter power spectrum $P_m$ through
\ba
\label{eq:Pgk}
P_h(k,M,z) &= \bigl[\bo^{\rm E}(k,M,z)\bigr]^2 P_m(k,z) \\
&= \bigl[b_1^{\rm E}(M,z) + \Delta \bo(k,M,z)\bigr]^2 P_m(k,z)
\nonumber\\
\label{eq:Dbk_fnl}
\Delta \bo(k,M,z)
&= \left(b_1^{\rm E}(M,z)-1\right)\frac{2\fnl\delta_c}{\Mm(k,z)} \;.
\ea
Here, $b_1^{\rm E}(M,z)$ is the linear, Gaussian (Eulerian) bias of halos
($b_1\equiv b_1^{\rm E}-1$ is the corresponding Lagrangian bias),
$\d_c\simeq 1.69$ is the critical density for spherical collapse, and 
the function $\Mm(k,z)\equiv 2k^2T(k)D(z)/(3\Omega_mH_0^2)$ depends on 
the linear growth factor $D(z)$, the matter transfer function $T(k)$, the 
present-day matter density parameter $\Omega_m$ and the Hubble constant 
$H_0$. In what follows, we shall omit the mass and redshift dependence for 
conciseness.  

The scale-dependent non-Gaussian (NG) bias expression [\refeq{Dbk_fnl}] may be 
extended (for $N=3$) using a peak-background split (PBS) argument \cite{fk10}. 
Alternatively, \refeq{Dbk_fnl} may be generalized to any model of primordial
NG by computing the correlation function of high threshold regions 
\cite{mv08}. Even though the result is strictly valid in the high-peak
limit only, i.e. $\nu \equiv \d_c/\s_{0s}\gg 1$ where $\s_{0s}$ is the 
variance of the density field smoothed on scale 
$R_s(M) = (3 M/4\pi \rhob)^{1/3}$, it is usually extended to arbitrary peak 
heights upon identifying $\nu/\s_{0s}$ with $b_1$. For a given arbitrary $N$, 
this gives
\be
\label{eq:Db_general0}
\Delta \bo^{\rm (hp)}(k)
= 
\frac{4 b_1 \d_c}{(N-1)!} 
\left(
\frac{\nu}{\sigma_{0s}}
\right)^{N-3}
\frac{{\cal F}_s^{(N)}(k)}{\Mm_s(k)} \;.
\ee
We have defined $\Mm_s(k) = \Mm(k)W_{R_s}(k)$, where $W_{R_s}(k)$ is the 
Fourier transform of the spherical tophat filter with radius $R_s$. 
Furthermore, we have introduced the shape factor
\ba
{\cal F}_s^{(N)}(k)
=\:&
\frac{1}{4\sigma_{0s}^2\,P_\phi(k)}
\left[
\prod_{i=1}^{N-2} \int\!\!\frac{d^3k_i}{(2\pi)^3}\, \Mm_s(k_i)
\right]
\Mm_s(q) 
\nonumber
\\
& \times 
\xi_\phi^{(N)}(\bm{k}_1,\cdots,\bm{k}_{N-2},\bm{q},\bm{k})\;,
\ea
with $\bm{q}\equiv-\bm{k}_1-\cdots-\bm{k}_{N-2}-\bm{k}$, and $\Pp(k)$ 
being the power spectrum of the Bardeen potential $\phi$. For the constant 
$\fnl\phi^2$ model, ${\cal F}_{s}^{(3)}(k)/\fnl$ is equivalent to the 
function ${\cal F}_R(k)$ defined in \cite{mv08}.
As ${\cal F}_s^{(3)}(k)\to \fnl$ on large scales, \refeq{Db_general0} 
reproduces \refeq{Dbk_fnl} in the limit $k\to 0$. Since, in the limit 
$\nu\gg 1$, both thresholding and PBS approaches yield \refeq{Db_general0}, 
we call it the \emph{high-peak} result.  

Measurements of the clustering of dark matter halos in non-Gaussian N-body 
simulations have confirmed the validity of \refeq{Pgk}. The $k$-dependent 
NG bias currently yields constraints on the nonlinear parameter $\fnl$ 
competitive with CMB bispectrum measurements \cite{ddhs08}. For other models 
of primordial non-Gaussianity however, comparisons between the high-peak expectation 
[\refeq{Db_general0}] and the simulated NG bias have revealed large 
discrepancies in the magnitude of $\D\bo$ \cite{ds10,sdh10,wv11}. 


In \cite{paper1}, we calculate the scale-dependent NG bias using two 
different formulations of the peak-background split. The first generalizes 
the approach adopted in \cite{ddhs08,fk10}; the second utilizes conditional 
mass functions. Both methods arrive at the same expression on large scales 
and thus validate the robustness of the result. Let us briefly review the 
second approach, which is inspired by a derivation of the Gaussian peak
bias factors \cite{dcss10}.  The conditional mass function 
$\bar{n}(M,R_s|\d_l,R_l)$ gives the mean number density of halos of mass 
$M$ and radius $R_s$ inside a large region of size $R_l\gg R_s$ and 
overdensity $\delta_l$. The conditional mass function encodes information 
on the halo bias parameters \cite{mw96,st99}. On taking the limit 
$R_l\to\infty$ and expanding the conditional over-abundance of halos, 
\be
\label{eq:dpk}
\left\langle \d_h|\d_l\right\rangle =
\frac{\bar{n}(M,R_s|\d_l,R_l)}{\bar{n}(M,R_s)}-1\;,
\ee
in powers of the large-scale perturbation $\d_l$, we are able to read off 
the NG bias factors. In particular, the scale-dependent NG contribution to 
$\bo^{\rm E}$ is at leading order
\ba
\label{eq:Db_general}
\Delta \bo^{\rm (pbs)}(k)
&= 
\frac{4}{(N-1)!} 
\frac{{\cal F}_s^{(N)}(k)}{\Mm_s(k)} \epsilon_s(k)
\\
\epsilon_s(k) &\equiv
b_{N-2}\d_c + b_{N-3}
\left(
3-N + \frac{d\ln{\cal F}_s^{(N)}(k)}{d\ln\sigma_{0s}}
\right)
\;,
\nonumber
\ea
where $b_N$ is the $N$-th order, Gaussian Lagrangian bias 
(with $b_0\equiv 1$).  
The first term ($b_{N-2}\delta_c$) reduces to \refeq{Db_general0} for high 
mass halos since $b_N\simeq (\nu/\sigma_{0s})^N$ when $\nu\gg 1$. Therefore, 
it exactly reproduces the previous results of \cite{ddhs08,ds10,fk10,sdh10}.   
The second term, which is derived for the first time in \cite{paper1}, only 
vanishes in the local, quadratic $\fnl$ model and on large scales. Although
it becomes sub-dominant in the high-peak limit ($b_{N-2}/b_{N-3}\gg 1$), it 
is significant for most relevant peak heights.

To decipher the origin of this new term, it is useful to recall the basic 
mechanism by which primordial non-Gaussianity affects halo clustering.
In the presence of a primordial bispectrum, the small-scale density field 
is modulated by a long-wavelength ($k \lesssim 0.01\iMpch$)
perturbation of the potential 
$\phi$ via \cite{ddhs08,fk10}
\be
\label{eq:hs0s_fnl}
\hat\sigma_{0s}^2(k) = \s_{0s}^2
\Bigl(1+4 {\cal F}_s^{(3)}\!(k)\phi_l(\vk)\Bigr)\;.
\ee
Thus, instead of being a constant throughout space, $\hat\sigma_{0s}^2$
and hence $\nu$ vary from point to point.  Since the local abundance
of halos depends on the amplitude of small-scale density fluctuations
(it is commonly considered to be proportional to some multiplicity function 
$f(\nu)$), the change in $\hat\sigma_{0s}$ [\refeq{hs0s_fnl}] contributes 
to the variation in the overdensity $\d_h$ of halos. Most importantly, this 
modulation of $\hat\s_{0s}$ induces a dependence of $\d_h$ on $\phi$, rather 
than $\d_m$.  This explains why the scale-dependent NG bias generally 
increases towards larger scales (but note the additional $k$-dependence 
brought by ${\cal F}_s^{(3)}(k)$). As shown in \cite{paper1}, similar 
arguments can be made for a generic primordial $N$-point function. In 
this case, long-wavelength potential perturbations locally induce a 
reduced moment of order $(N-1)$ in the small-scale density field $\d_s$, 
which is given by 
\be
\hat {\cal S}_s^{(N-1)}(k) \equiv \frac{\<\d_s^{N-1}\>}{\s_{0s}^{N-1}}
= 4 \s_{0s}^{3-N} {\cal F}_s^{(N)}(k) \phi_l(\vk)\;.
\ee
For example, a primordial trispectrum generates a local skewness.  
Therefore, we can think of the effect of a primordial $N$-point function as 
locally rescaling the local significance according to \cite{paper1}
\begin{equation}
\label{eq:hatnu}
\nu \to \hat\nu \equiv \nu
\biggl(1 - \frac{4}{(N-1)!}b_{N-3} {\cal F}_s^{(N)}(k)\phi_l(\vk)
\biggr)\;.
\end{equation}
Clearly, the modulation of the variance, skewness etc. of the small-scale 
density field not only changes the significance $\nu$ that corresponds to a 
mass $M$, but also affects the significance interval $d\nu$ which a fixed 
mass bin $dM$ is mapped to.  
Hence, long-wavelength perturbations of the Jacobian $d\ln\hat\nu/d\ln M$ 
generate a dependence of the non-Gaussian halo bias on the derivative of 
the shape factor ${\cal F}_s^{(3)}$ w.r.t. $R_s$. This term must be present 
because halos in N-body simulations and, to a lesser extent, in galaxy 
surveys are identified by mass.  
In contrast, in the thresholding approach the two-point correlation function
of halos above a mass threshold $M$ is associated with that of Lagrangian 
regions above a threshold $\d_c$ in the linear density field smoothed on a
{\it fixed} scale $R_s(M)$. This is the reason why this approach cannot 
recover the second term in the expression of $\epsilon_s(k)$ 
[\refeq{Db_general}].

In the remainder of this Letter, we shall compare the prediction \refeq{Db_general} to 
simulation results in the large-scale limit $k\to 0$. In this regard, it 
is convenient to express both the predicted and simulated scale-dependent 
NG bias in units of the high-peak expectation \refeq{Db_general0}, i.e. we 
plot $\D\bo/\D\bo^{\rm hp}$. Our prediction for this ratio is
\be
\label{eq:ratio}
\frac{\Delta \bo^{\rm (pbs)}(k)}{\Delta \bo^{\rm (hp)}(k)} = 
\left(\frac{\nu}{\s_{0s}}\right)^{3-N} \frac{\epsilon_s(k)}{b_1 \d_c}
\ee
for a given primordial $N$-point correlation function. Note that, in the 
models we consider below, $\epsilon_s$ depends very weakly on $k$ for 
wavenumbers $k\lesssim 0.01 \iMpch$. Hence, \refeq{ratio} predicts a
 mass-dependent correction to the amplitude of the scale-dependent NG bias 
on large scales.
  
\begin{figure}
\includegraphics[width=3.2in]{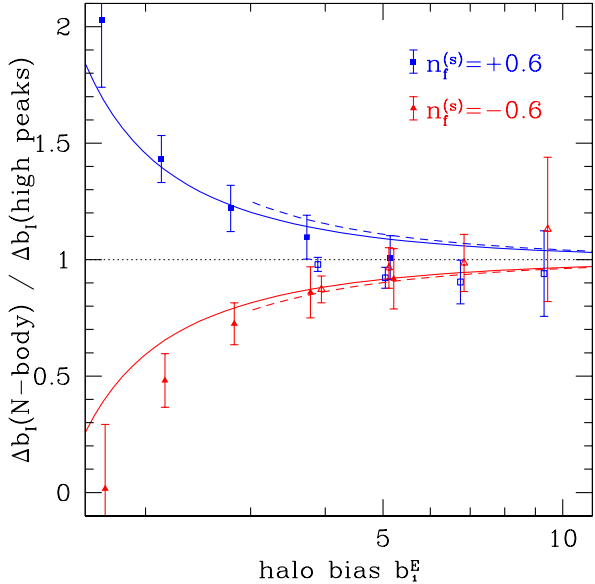}
\caption{
Ratio of the large scale NG halo bias measured in simulations with 
non-Gaussianity of the scale-dependent local type ($n_f=\pm 0.6$) 
\cite{sdh10}, to that predicted by the high-peak approximation.  
Filled and open symbols show the simulation results, whereas the 
solid and dashed curves indicate the new theoretical expectation
[\refeq{ratio}] at $z=0$ and 1, respectively.} 
\label{fig:fig1}
\end{figure}

We consider the following three models beyond the scale-independent 
local $\fnl$ for which simulations have been performed:
\begin{itemize}
\item{scale-dependent $\fnl$ \cite{sdh10}
\be
\xi_\phi^{(3)}(\vk_1,\vk_2,\vk_3)
=\fnl(k_p)\left(\frac{k_1}{k_p}\right)^{n_f} P_2 P_3
+(5~\mathrm{perm}.)
\label{eq:xi3kfnl}
\ee
}
\item{orthogonal bispectrum \cite{wv11}
\ba
\xi_\phi^{(3)}(\vk_1,\vk_2,\vk_3) &=
6\fnl\big[P_1 P_2+(2~{\rm perm}.) + 3(P_1 P_2 P_3)^{2/3} 
\nonumber \\
& \quad -(P_1^{1/3} P_2^{2/3} P_3 +(5~{\rm perm}.))\big]
\label{eq:xi3ortho}
\ea
}
\item{local $\gnl$ model \cite{ds10}
\be
\xi_\phi^{(4)}(\vk_1,\vk_2,\vk_3,\vk_4) =
6\gnl\bigl[P_1 P_2 P_3 + (3~{\rm perm}.) \bigr]\;,
\label{eq:xi4gnl}
\ee
}
\end{itemize}
where $P_i\equiv \Pp(k_i)$.  The shape factor for each of these models
on large scales respectively asymptotes to
\ba
\label{eq:kfnl_lowk}
{\cal F}_s^{(3)}(k) &=
\fnl(k_p) k_p^{-n_f} \left(\frac{\s_{\a s}}{\s_{0s}}\right)^2\;, 
\   \a=n_f/2 \\
\label{eq:ortho_lowk}
{\cal F}_s^{(3)}(k) &= 
-3\fnl \left(\frac{\s_{\a s}}{\s_{0s}}\right)^2 k^{-2\alpha}\;,
\   \a=(n_s-4)/6 \\
\label{eq:gnl_lowk}
{\cal F}_s^{(4)}(k) &=
\frac{3}{4}\gnl \s_{0s}^2\Sk_{s,\rm loc} \;, 
\ea
where the spectral moments are defined as
\be
\sigma_{n s}^2\equiv
\int\!\!\frac{d^3k}{(2\pi)^3}\, k^{2n} \Pp(k) \Mm_s^2(k)\;.
\ee
In \refeq{gnl_lowk}, $\Sk_{s,\rm loc}$ is the skewness of the density field
smoothed on scale $R_s$ in a local quadratic model with $\fnl=1$.  Using 
these expressions, it is straightforward to evaluate the ratio \refeq{ratio}.  

\reffig{fig1} shows the simulations results of \cite{sdh10} for the local, 
scale-dependent $\fnl$-type model [\refeq{xi3kfnl}] relative to the high-peak 
prediction, as a function of the Gaussian halo bias $b_1^{\rm E}$. 
The values of $k_p=0.04$~Mpc$^{-1}$ and $\fnl(k_p)=630$ are fixed, while 
the spectral index is $n_f=\pm 0.6$. Clearly, the data points are 
inconsistent with the high-peak prediction in the range 
$b_1^{\rm E}\lesssim 4$.  
The improved theoretical prediction presented in this paper explains the 
observed deviation of the high-peak prediction for the $z=0$ case (filled 
symbols, solid lines) and  for the $z=1$ case and negative $n_f$ (open 
symbols, dashed lines).  
The only exception is for $n_f=0.6$ at $z=1$, where the N-body data appears 
consistent with the high-peak prediction. This case deserves further 
investigation.  Note however that the simulation measurements are
correlated, since they were estimated from a common simulation volume.  
Hence, the significance of this departure is not straightforward to estimate.

\begin{figure}
\includegraphics[width=3.2in]{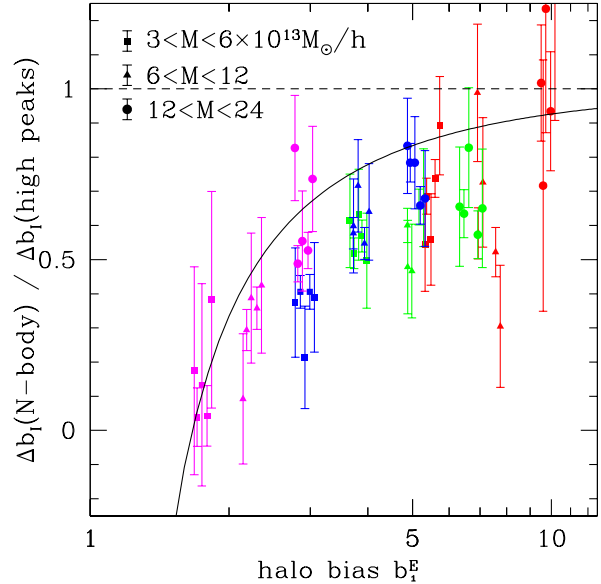}
\caption{
Same as \reffig{fig1} but for the orthogonal bispectrum shape \cite{wv11}. 
Different symbols correspond to various halo mass bins as indicated in the 
figure; colors indicate the redshifts $z=0$ (magenta), 0.67 (blue), 1 
(green) and 1.5 (red).  The solid curve shows the new theoretical 
prediction assuming a halo mass $M=5\times 10^{13}\hmsun$.
}
\label{fig:fig2}
\end{figure}

The results from simulations of the orthogonal bispectrum [\refeq{xi3ortho}] 
are shown in \reffig{fig2} \cite{wv11}. Symbols represent the measured NG 
halo bias from simulations relative to the high-peak prediction from three 
realizations with $\fnl=-250$ and two with $\fnl=-1000$. The results were 
averaged over wavenumbers with $k<0.1\hmmpc$. 
We show our prediction (\refeq{ratio} with $N=3$ together with 
\refeq{ortho_lowk}) as the solid curve, for a median halo mass 
$M=5\times 10^{13}\hmsun$. Again, the improved theory is consistent 
with the N-body data, which convincingly shows a strong suppression for 
$b_1^{\rm E}\lesssim 2$.  While the error bars are significant,
the strong mass- and redshift-dependence predicted by \refeq{Db_general} 
appears to be supported by the data.

\begin{figure}
\includegraphics[width=3.2in]{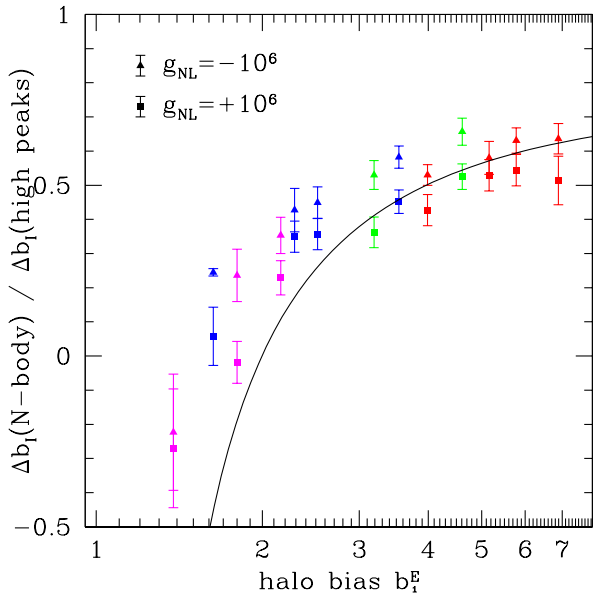}
\caption{Same as \reffig{fig1} but for the local cubic model. Triangle 
and square symbols indicate results for $\gnl=-10^6$ and $10^6$, 
respectively. Different colors show different redshifts spanning the 
range $0<z<2$.}
\label{fig:fig3}
\end{figure}

Finally, \reffig{fig3} shows the ratio of the simulated large-scale bias of 
the local $\gnl$ model [\refeq{xi4gnl}] to the high-peak prediction. 
The simulations used values of $\gnl=\pm 10^6$ \cite{ds10}.  Error bars 
indicate the scatter among 5 realizations.  Clearly, the measured values of 
$\D\bo$ lie far below the high-peak prediction.  While they increase 
monotonically with $b_1^{\rm E}$, $\D\bo/\D\bo^{\rm (hp)}$ never reaches 
unity even for the most biased samples.  The solid curve represents the 
improved theoretical prediction, \refeq{ratio} with the shape factor 
\refeq{gnl_lowk}.  For this Figure, we have 
computed the Gaussian peak-background split biases $b_1$ and $b_2$ from a 
Sheth-Tormen mass function \cite{st99}.  The prediction matches the 
measurements well, though it somewhat underestimates $\D\bo$ for 
$b_1^{\rm E}\lesssim 2$;  an improved match might be achieved by
measuring the $b_2$ of halos directly from simulations, rather than
using the fitting function of \cite{st99}.  
Note also that the measurements are noticeably larger for $\gnl=-10^6$, 
suggesting that second order contributions in $\gnl$ are 
important. This is presumably due to the fact that the Gaussian bias
factors $b_1$ and $b_2$ receive scale-independent NG corrections which 
depend on the nonlinear parameter $\gnl$. 


In summary, we have presented a new formula for the NG bias 
based on an improved peak-background split argument \cite{paper1}.  
We have obtained an additional term which can be interpreted as the 
effect of non-Gaussianity on the mapping between the significance 
$\nu$ and the mass of halos $M$.  This new term vanishes only for the 
local quadratic model with constant $\fnl$, and is significant for all 
other models considered here.  We compare our theoretical predictions 
to the scale-dependent bias measured from NG N-body simulations, and 
show that it is in good overall agreement with the measurements, in
contrast to the existing formulae based on the statistics of high peaks.  
The improved theoretical predictions presented here will enable 
accurate upper limits on, or measurements of, general shapes of 
non-Gaussianity.  In particular, our results are of relevance to
the recent measurements of \cite{XiaEtal11}, who employed the 
high-peak prediction, and are expected to increase their upper limits 
by factors of order unity.  
We expect similar mass-dependent corrections to the quadratic and 
higher-order bias factors which should significantly affect the halo 
bispectrum \cite{paper1}.  
Accounting for all these corrections will be essential to obtain accurate 
constraints on primordial non-Gaussianity from large-scale structure.  

We would like to thank Sarah Shandera and Christian Wagner for providing 
the data shown in \reffig{fig1} and \reffig{fig2}, respectively. DJ
and FS are supported by the Gordon and Betty Moore Foundation at Caltech. 
VD is supported by the Swiss National Foundation under contract No. 
200021-116696/1 and FK UZH 57184001.

\bibliographystyle{prsty}

\end{document}